\documentclass[sigconf]{acmart}

\usepackage{graphicx}
\usepackage{subcaption}

\usepackage{ifpdf}
\ifpdf
\copyrightyear{2026}
\acmYear{2026}
\setcopyright{cc}
\setcctype{by}
\acmConference[UIST '26]{The 39th Annual ACM Symposium on User Interface Software and Technology}{November 02--05, 2026}{Detroit, MI, USA}
\acmBooktitle{The 39th Annual ACM Symposium on User Interface Software and Technology (UIST '26), November 02--05, 2026, Detroit, MI, USA}
\acmDOI{10.1145/3830398.3830639}
\acmISBN{979-8-4007-2856-3/2026/11}
\fi

\begin{document}

\title{Social.Wiki: A Web Held in Common}

\author{Theia Henderson}
\affiliation{
  \institution{MIT CSAIL}
  \city{Cambridge}
  \state{MA}
  \country{USA}
}
\email{theia@mit.edu}
\author{Carmel Schare}
\affiliation{
  \institution{MIT CSAIL}
  \city{Cambridge}
  \state{MA}
  \country{USA}
}
\email{schare@mit.edu}
\author{Ana Dodik}
\affiliation{
  \institution{MIT CSAIL}
  \city{Cambridge}
  \state{MA}
  \country{USA}
}
\email{anadodik@mit.edu}
\author{Clemens N. Klokmose}
\affiliation{
  \institution{Aarhus University}
  \city{Aarhus}
  \country{Denmark}
}
\email{clemens@cs.au.dk}
\author{Ziv Epstein}
\affiliation{
  \institution{MIT College of Computing}
  \city{Cambridge}
  \state{MA}
  \country{USA}
}
\email{zive@mit.edu}
\author{David D. Clark}
\affiliation{
  \institution{MIT CSAIL}
  \city{Cambridge}
  \state{MA}
  \country{USA}
}
\email{ddc@csail.mit.edu}
\author{David R. Karger}
\affiliation{
  \institution{MIT CSAIL}
  \city{Cambridge}
  \state{MA}
  \country{USA}
}
\email{karger@mit.edu}

\begin{teaserfigure}
  \includegraphics[width=\textwidth]{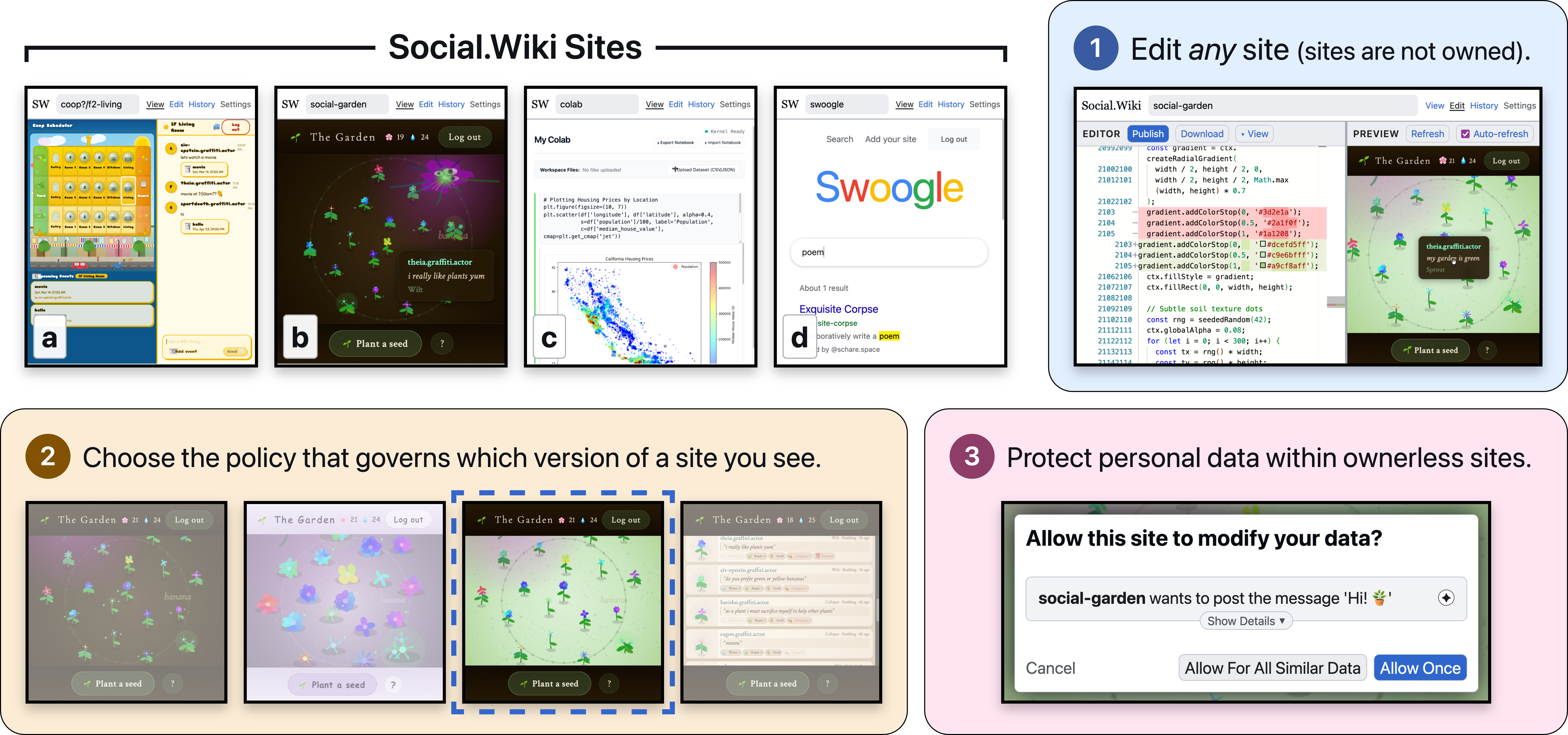}
  \caption{An overview of Social.Wiki, a reimagining of the web where sites are collaboratively edited like Wikipedia articles. Examples of sites in the top left were built by participants in two hours at a Social.Wiki hackathon: (a) a messenger and event scheduler for a housing co-op; (b) a creative microblogging site where posts are visualized as flowers; (c) a collaborative Python notebook; (d) a socially constructed web directory.
  }
 \Description[Example Social.Wiki sites plus demonstration of editing, plural governance, and security]{%
    The figure consists of four components: a collection of "Social.Wiki Sites", plus 3 numbered panels, 1 through 3, detailing properties of those sites. The Social.Wiki Sites are 4 screenshots labeled a through d. Each has the same bar at the top that says "SW" followed by a navigation bar with text in it and four buttons: View, Edit, History, and Settings. The View button in each is selected.
    Site a.
    The text in the navigation bar says "coop?/f2-living". The app below the header is labeled "Coop Scheduler" and is divided into two panels. The left panel contains a cartoon house with different rooms. "Living" on the second floor is selected. There is also a calendar of upcoming events. On the right, there is a chat interface. The title says "2F living room" and there are messages below, some of which refer to events. The chat submission button has an option to "Add event".
    Site b.
    The text in the navigation bar says "social-garden". The app below the header is dirt-colored with a header, footer, and main canvas. The header says "The Garden", and then has numeric metrics labeled with a flower emoji (19) and water emoji (24). The footer says "Plant a seed" with an informational question mark next to it. The canvas contains a group of cartoon flowers. One is hovered over, displaying the username "theia.graffiti.actor" and text "i really like plants yum".
    Site c.
    The text in the navigation bar says "colab". The app below the header is a notebook interface titled "My Colab". It contains Python code for "Plotting Housing Prices By Location" and includes a colorful visualization of "California Housing Prices".
    Site d.
    The text in the navigation bar says "swoogle". The app below the header is a search interface with the title "Swoogle" in colorful font. At the top, there is an option to "Add your site". The search bar has the text "poem" entered. Below is "About 1 result" labeled "Exquisite corpse" with a description "Collaboratively write a poem". The text "poem" is highlighted.
    Panel 1.
    Panel 1 is labeled "Edit any site (sites are not owned)". The screenshot below has the same header as the Social.Wiki sites (SW, navigation bar, four buttons: View, Edit, History, Settings). However, in this panel "Edit" is selected. The navigation bar includes the text "social-garden". Below are two panes: "Editor" and "Preview". The "Editor" shows a code diff of changing hex colors and has options "Publish" and "Download". The "Preview" shows an app similar to site b: "The Garden" with an option to "Plant a Seed" and a canvas of flowers. Rather than the background of the canvas being dirt-colored, it is light green. The text box of a highlighted flower reads "my garden is green".
    Panel 2.
    Panel 2 is labeled "Choose the policy that governs which version of a site you see." There are four variations of the "social-garden" app from site b and panel 1. The third variant has a selection rectangle around it and the others are greyed-out. The first variant is the same as in site b: a header that says "The Garden" a footer that says "Plant a Seed" and a dirt colored background. The second variant has a light purple-colored background, uses comic-sans font, and shows the flowers enlarged with no stems. The third variant is the same as in panel 1 which is identical to variant 1 but has a light green colored background rather than a dirt colored background. The fourth variant has the same header and footer as variants 1 and 3 but in the middle displays the flowers as posts in feed. A flower is on the left and post text is on the right with a username.
    Panel 3.
    Panel 3 is labeled "Securely interact with changing, ownerless sites by granting data access as needed." Below is a screenshot of a popup with the title "Allow this site to modify your data?" It has the options "Cancel", "Allow For All Similar Data", and "Allow Once". In the middle is a box with the text "social-garden wants to post the message Hi!". The middle text box has an AI sparkle on the right and an option to "Show Details" with an arrow pointing down.
  }
  \label{fig:teaser}
\end{teaserfigure}

\begin{abstract}
Many of the websites people depend on have owners whose interests are not fully aligned with their users.
We address the root of this problem by presenting a reimagining of the web where sites are \emph{not owned at all} but are instead collaboratively produced like Wikipedia articles.
We call the system Social.Wiki because it supports the co-creation of interactive \emph{social} sites, such as those for microblogging, messaging, dating, gaming, ride sharing, and so on.
With off-the-shelf AI tools, people with little or no programming experience can edit these sites to better reflect the needs and preferences of their communities.

Social.Wiki builds on ideas from collaborative malleable software systems such as Webstrates, but is designed for public participation rather than use only within small, trusted groups.
To this end, Social.Wiki includes governance to mitigate conflict.
To accommodate diverse governance preferences,
our model of ``\emph{plural governance}'' lets people independently choose the policies that determine which edits to a site they see.
Social.Wiki also implements a granular security model to protect personal data in a malleable environment.

Complementing the decentralized design and governance of Social.Wiki sites, both site edits and within-site data are stored on Graffiti, a decentralized infrastructure, decoupling the ownership of underlying servers from the ownership of sites.

We evaluate Social.Wiki through case studies that demonstrate the range of sociotechnical structures it supports, as well as through deployments at a hackathon and in the wild.
\end{abstract}


\maketitle
\section{Introduction}

Sites on the World Wide Web have \emph{owners}.
At first, this design decision might seem inevitable,
both as a technical means of delegating the ``server'' of a particular site
and as a social means of governing the web.
However, it is arguably because of this ownership model that most people are mere tenants
within the sites they must inhabit to live their lives---subject to impersonal and often hostile designs simply to coordinate carpools, advertise their businesses, or connect with friends.

This work offers a reimagining of the web where all sites are instead \emph{held in common}.
On this web, any visitor to a site may also participate in the design of that site itself.
This not only gives people the freedom to remove harmful ``dark patterns,''
but also allows communities to evolve the sites they rely on into spaces that are highly personalized to their needs, preferences, and values.

Our prototype of this web-in-common is called Social.Wiki.
Its client interface, shown in Figure~\ref{fig:teaser}, combines elements of both a web browser and Wikipedia.
Like the web, one can browse Social.Wiki to find all sorts of sites, including, importantly,  \emph{social} sites such as those for messaging, microblogging, meeting neighbors, dating, playing games, and so on.
Like Wikipedia, every Social.Wiki site has an ``Edit'' button that any visitor can use to modify that site's content, styling, and functionality.

If Social.Wiki were implemented like a traditional wiki, with our central server storing the edits, then it would be making the problem worse: exchanging the owners of individual sites with ourselves as the owners of \emph{all} sites.
Instead, Social.Wiki is implemented on top of Graffiti~\cite{graffiti}, a decentralized social infrastructure.
Effectively, each edit to a Social.Wiki site produces a new version, stored on a server of the \emph{editor's choice}.
When someone visits a Social.Wiki site, their client retrieves information about these versions from the various chosen servers and selects one version to download and present.
Thus, just as no one owns the web in its entirety, no one owns Social.Wiki.
Admittedly, to provide a convenient entry point into Social.Wiki, our reference Social.Wiki browser is published on the (world wide) web at a domain that we do own\footnote{\url{https://social.wiki}}.
However, this browser is open source\footnote{\url{https://github.com/sportdeath/socialwiki}} and other competing browsers can be built and distributed, just like alternative web browsers.

A system where anyone can edit any site immediately raises the potential of destructive vandalism and conflict. 
The typical response in ``peer production''~\cite{peerproduction2} systems is to introduce a policy that governs changes to the editable artifacts.
However, this work is motivated by the critique of one such policy---ownership. So, rather than replacing one normative means of governing an entire web with another, Social.Wiki
draws on pluralist views of design and governance~\cite{escobarpluriverse,designjustice,feministhci,beyondpanaceas,modularpolitics}
to introduce what we call ``\emph{plural governance},'' which allows for multiple, possibly conflicting governance policies to coexist.
Different coexisting governance policies may select different versions of a site to present. 

Whichever governance policies people choose, no policy can catch every harmful edit to a site.
Therefore, we also present security safeguards that protect against the damage that a malicious (or simply buggy) site version could otherwise inflict by destroying, leaking, or forging personal data. 

By providing governance and security,
Social.Wiki extends ``computational media'' to public participation settings. Computational media is a model of collaborative, malleable software where the boundary between user and developer is blurred~\cite{computationalMedia}. Prior systems, such as Webstrates~\cite{webstrates}, rely on self-proclaimed ``toilet door security'' that does not scale beyond small, trusted groups~\cite{computationalMedia}.

\subsection{Governance Overview}

Like the web, each Social.Wiki site may contain \emph{its own} governance for within-site interactions: a microblogging site may have a mechanism for flagging posts, and a messaging site may have a mechanism for blocking message requests.
The governance that Social.Wiki provides is a layer above: it regulates the editing of \emph{sites themselves}---including the editing of a site's internal governance.

By default, the Social.Wiki browser provides a governance policy inspired by Wikipedia's battle-tested system, but plural governance lets people freely replace it for themselves.
Governance can be replaced browser-wide, or sites can act as localized governance ``lenses'' that provide views into alternative governance systems. 
For example,
the site ``Mypage,'' described in Section~\ref{sec:mypage}, reproduces the governance of ownership by filtering a site's edits for those by a single person (its ``owner''). Mypage provides a space for people to create and view personal homepages but without imposing ownership on the system as a whole.

\subsection{Personal Data Ownership}
\label{sec:personaldata}

This work argues for eliminating the ownership of \emph{shared social spaces}.
Ownership of such spaces gives owners power over participants, including the power to determine who may participate and under what conditions.
However, we actively work to protect the ownership of personal data because this form of ownership establishes an opposite social relation: rather than granting power over others, it protects an individual's agency within a shared space.

In the traditional web model, the responsibility of protecting personal data is delegated to the owners of the sites that the data is produced on.
The outcome is often disappointing, and in Social.Wiki, there are no owners to whom we could delegate such responsibility, even if we wanted to.
Instead, all of a user's data is stored on Graffiti (again, on the server of their choice) and all changes to that data are mediated by a client-side ``\emph{data guard}.''
Just as a web browser mediates access to a camera or microphone, the guard confirms new data operations with the user before executing them---``Allow this site to modify your data?'' 
This mediation prevents Social.Wiki's openly editable sites from silently accessing, creating, or deleting personal data, whether through malicious intent or, more innocently, due to bugs from inexperience or ``vibe coding.''

\section{Motivation}
\label{sec:motivation}

This work is motivated by problems arising from the private
ownership of a certain class of websites that represent shared social spaces, often called \emph{platforms}~\cite{srnicek2016platform}.
Although Social.Wiki
is a general system for collaboratively producing \emph{many} kinds of sites, it notably supports the co-production of platforms.
As such, it demonstrates a new model for
how platforms might be
created and governed without private ownership. 
This section reviews the problems associated with privately owned platforms and situates Social.Wiki within the broader landscape of proposed solutions.

Per \citet{srnicek2016platform}, we consider platforms to be infrastructures that facilitate digital connections between people.
These platforms can be used for
socializing, entertainment, conducting transactions, and organizing workplaces,
to name a few.
Examples include Facebook, Discord, Tinder, Reddit, Uber, Etsy, and Spotify.
Owning a platform implies the ``ownership of software (the 2 billion lines of code for Google,  or the 20 million lines of code for Facebook) and hardware (servers, data centres, smartphones, etc.)''~\cite{srnicek2016platform}, as well as the data extracted from users~\cite{couldry2019colonialism}.
The default ownership structure of today's web is private ownership, which puts the power
to design platforms
into the hands of private individuals or shareholders.
Notably, this leaves most users out of the design process.

The power asymmetry between platform owners and users can lead to harms.
In some cases, owner decisions raise macro-scale concerns about user privacy~\cite{mejias2024data, couldry2019colonialism}, the commodification of sociality~\cite{bollmer2024influencer,jarrett2014womens,couldry2019costs}, or disproportionate harms to minoritized communities~\cite{platformedracism,disproportionateremovals}.
Other times, decisions can manifest as deceptive ``dark patterns''~\cite{darkpatternsatscale,darkpatternsux}
or labor exploitation~\cite{algorithmicdomination,levy2023truckers,muldoon2024feeding}.
Recent years have seen an increase in these trends due to network effects~\cite{platformcompetition,srnicek2016platform} that entrench dominant platforms,
a dynamic described as platform decay~\cite{platformdecay} or more popularly ``enshittification''~\cite{enshittification,enshittificationbook}.

Many proposed solutions to platform harms involve the creation of structures that encourage platform owners to be more accountable to their users.
Sometimes, these structures are self-imposed, by incorporating the platform as a nonprofit~\cite{dpi}, or building it upon an open protocol such as ActivityPub~\cite{activitypub}, the AT Protocol~\cite{atprotocol}, or Graffiti~\cite{graffiti}, each of which makes it easier to leave one platform for another.
Alternatively, these structures can be externally imposed via legislation, such as the Digital Markets Act~\cite{dma}.

While owner accountability measures may mitigate the most egregious harms, and there are cases of long-term success (\emph{e.g.}, Vermont's Front Porch Forum, Archive of Our Own~\cite{threeleggedstool}), they leave the underlying power distribution intact.
The owners still retain the ultimate control over the design and governance of the platform,
while users must rely on those owners to remain responsive despite changing leadership, incentives, or market pressures that can drive platform decay~\cite{papadimitropoulos2021coop}. 
Users retain the ability to leave---made easier by open protocols that let them keep their data~\cite{activitypub,atprotocol,graffiti}---and can influence the platform by threatening to do so~\cite{hirshman2009hirshman}.
However, leaving still means abandoning the existing system rather than improving it.
In contrast, we are interested in enabling users to \emph{collectively} work to construct better platforms.

Platform cooperativism is an alternative model where a platform is collectively owned by its participants~\cite{scholz2016platform,papadimitropoulos2021coop}.
However, platform cooperatives are difficult to produce because they bootstrap on the legal system; the first author knows from firsthand experience how difficult it is to create a legal cooperative in the United States.

Instead of having to rely on a patchwork of legal frameworks, Social.Wiki can be used to create new platforms that are collectively designed and governed by \emph{default}.
Ultimately, any Social.Wiki site \emph{could} be constructed by building on an open protocol and crafting a bespoke legal solution with commons-based licenses and co-ownership guarantees.
Our system unifies these demands into one system that supports the creation of cooperative platforms through \emph{technical} means (decentralized hosting, production, and governance) rather than through legal means.

In some ways, these technical changes actually \emph{simplify} the process of making a website by removing the need to purchase and configure a domain name or deploy a server. This ease was noted by participants in our hackathon (\S\ref{sec:hackathon}).
This makes it possible to build experimental new platforms ``cooperative-first'' rather than  cooperativism being a reaction to already established power.

Some platforms cannot exist efficiently in Social.Wiki. In particular, aside from the ability to query a collective database, all computation happens on the \emph{client}. This is fine for platforms that are scoped in some way (follow-based microblogging, location-based services, group chats) but not for algorithmic feeds or recommendation systems that must consume vast amounts of data.
This limitation is inherited from the Graffiti system on which we build, and future workarounds are discussed by \citet{graffiti}.
\section{Related Work}
\label{sec:related-work}

\subsection{Peer Production}
\label{sec:peerproduction}

Peer production is a mode of creating goods and services through self-organizing communities rather than centralized control~\cite{peerproduction}. It is most prominently exemplified by free and open source software (FOSS) and Wikipedia, both of which demonstrate that loosely coordinated contributors can produce complex artifacts at scale.

This work builds off of the wiki model of peer production which provides a relatively simple means of direct editing. While FOSS projects include more expressive version management tools, like git, these tools can be complicated for novices~\cite{de2016purposes}.

Most peer production systems inevitably involve governance to maintain standards and resolve conflicts.
While some are ruled by a ``benevolent dictator for life,''~\cite{governablespaces}, many adopt democratic practices.
However, no governance system is perfect (a point equally true in the physical world).
After all, Wikipedia's governance, while arguably the most successful, is still heavily debated 25 years on~\cite{Halfaker_2012, Ajmani_2023, Car_2025, Wagner_2021}.
For a system like Social.Wiki---which does not have the unified mission of creating an encyclopedia, but instead aims to house a myriad of different experiences and communities---we expect there to be even more disagreement.

Alternatively, remixing is a form of peer production that avoids the problem of governance because editors create independent derivatives rather than contribute to shared artifacts.
To our knowledge, all systems for direct editing social apps use a remix system, including Scratch~\cite{resnick2009scratch} and Websim~\cite{vorkel2024websim}, avoiding the question of choosing one system to govern a diversity of social apps.
However, remix has the drawback of diffusing collaboration between non-canonical artifacts: on Scratch, only 3\% of remixes build upon another remix rather than building on a fresh project~\cite{hill2010responses}, whereas on Wikipedia, the median article 
passes through 14 rounds of editing by different contributors, and highly visible articles undergo hundreds of rounds of editing~\cite{kimmons2011understanding}.

With plural governance, this work attempts to strike a middle ground that does not diffuse collaborative energy but at the same time does not impose a fixed governance model on all participants.

\subsection{Malleable Software}
\label{sec:malleable}

Malleable software is software that can be reshaped with minimal friction to suit individual needs \cite{litt2025malleable}.
Canonical historical examples are Smalltalk \cite{kay1996smalltalk} and HyperCard \cite{goodman1998complete}, but spreadsheets are arguably the most durable example of malleable software currently used by the general public.

Editing malleable software has always been a collaborative practice, but in the early days this was achieved by trading floppy disks.
More recently, malleable software systems offer the
ability to collaboratively edit software in realtime over the internet.
This trait was introduced by Croquet~\cite{smith2003croquet}, a Smalltalk-based collaborative virtual 3D world, then popularized by Google Sheets, and generalized by Webstrates~\cite{webstrates}, a system that persists and shares changes to web pages in realtime.
As the web platform is quite powerful, Webstrates makes it possible to collaborate on software for document editing, games, video conferencing, and so on~\cite{webstrates,codestrates,mirrorverse,spatialstrates}.
A decentralized implementation, called MyWebstrates, lets people collaborate without a centralized server~\cite{mywebstrates}, and the related system Patchwork introduces git-like versioning~\cite{patchwork}.

Realtime collaboration in a computational environment blurs the distinction between data and code.
For example, in a Webstrates chat app 
there is no difference between sending a message and adding the message to the app's source code.
While this model is elegant,
it obscures the semantics of what the change is for: is it a routine message, or is it a deeper change to the app's functionality?
Notably, this makes it impossible to release a Webstrate on the open internet without either making it read-only---disabling both malleability \emph{and} social interactivity---or making it publicly writable and therefore vulnerable to abuse.

Social.Wiki extends the Webstrates vision of collaboratively editable websites by introducing governance and security properties that aim to make it feasible to widely publicize sites while maintaining malleability and social interactivity.
Doing so involves a fundamental change to the architecture that (re)introduces a separation of code and data.
Code is used to describe a site's interface and functionality, whereas data, representing within-site user activity, is populated into the site during runtime via queries.
Changes to the code can be \emph{governed}
without hindering regular user interactions like sending messages.
Meanwhile, changes to data are made \emph{secure}
against possibly malicious sites that could otherwise delete important data or post under a user's identity without permission.

Additionally, the separation of code and data extends Webstrates by making it possible for the data populated into a site to be specific to the person viewing it.
For example, the messaging app we discuss in Section~\ref{sec:ourchats} displays only the conversations that the current user is part of, rather than \emph{all} conversations created within the site---this is crucial for preventing disclosure of private conversations.
Separating code and data also makes it possible to build apps that present slices of very large datasets, like the ride sharing app we describe in Section~\ref{sec:rideshare} or Social.Wiki itself, which is built recursively in its own malleable system.

The elephant in the room with malleable software is whether people have the skills, time, or courage to change software.
This is arguably why spreadsheets maintain popularity as they have a much lower barrier to entry despite a lower ceiling of expressivity.
However, AI is rapidly changing this picture.
Many people who don't consider themselves programmers now engage in ``vibe coding''~\cite{pierce2026personal_software, roose2025vibecoding}, and a host of tools are being developed to streamline the process~\cite{yining2025generative, min2025malleable}.
With Social.Wiki, AI is not embedded in the system itself,
but we have created a Social.Wiki site that
provides a prompt that can be used with any AI chatbot or agentic coding tool to edit or generate Social.Wiki sites with little to no experience. This tool was used extensively in our Social.Wiki deployment (\S\ref{sec:deployment}).
\section{System Design}
\label{sec:default}

On the surface, Social.Wiki is a Wikipedia-like system for editing interactive websites rather than static articles.
One can browse Social.Wiki either by entering the ``name'' of a site in the navigation bar or by clicking on a link to the site.
Every site can be interacted with through three tabs: ``View,'' which simply presents the site, ``Edit,'' which allows one to modify the site as a document and publish new versions, and ``History,'' which allows one to navigate through different published versions of the site.
The system is governed by a Wikipedia-like model where certain users called ``trusted editors'' can mark sites as protected, thereby limiting who has edit access.

Under the surface, the system is more abstract: View, Edit, and History are all \emph{themselves} editable documents that we call ``governance lenses.''
Modifying the governance lenses changes how Social.Wiki sites are governed by changing which site versions are selected for display and how those versions are created and annotated.
To avoid having to recursively govern the governance lenses, modifications to these lenses are only applied locally.

The root of the system is a browser.
It provides Social.Wiki's top-level navigation bar, a default set of governance lenses,
and a simple interface for modifying the governance lenses (the Edit lens does not edit itself).
The browser also establishes a connection to a database. The governance lenses use this database to publish and fetch site versions while sites themselves might use the database to publish and fetch social data like posts, messages, and reactions.
We use the Graffiti system as the database because it is decentralized and provides a minimal-but-expressive client-side API that makes it possible to build a wide range of apps~\cite{graffiti}.

Different versions of the same site ``interoperate'' because they draw their social data from the same underlying database. Therefore, competing governance lenses can present different site versions without isolating their users from one another: posts on one site version can be read on another. The coexistence of different governance lenses is what we call ``plural governance.''

The connection to the database is mediated by a client-side ``data guard,'' loaded by the browser, that surfaces access requests to the user so that users can securely browse collaboratively produced sites without those sites being able to destroy, leak, or forge user data.
Additionally, every layer of ``transclusion''---including a document within another document~\cite{xanadu,webstrates}---is sandboxed so that sites cannot manipulate the governance lenses or the browser.

\begin{figure}[t]
    \centering
    \includegraphics[width=\columnwidth]{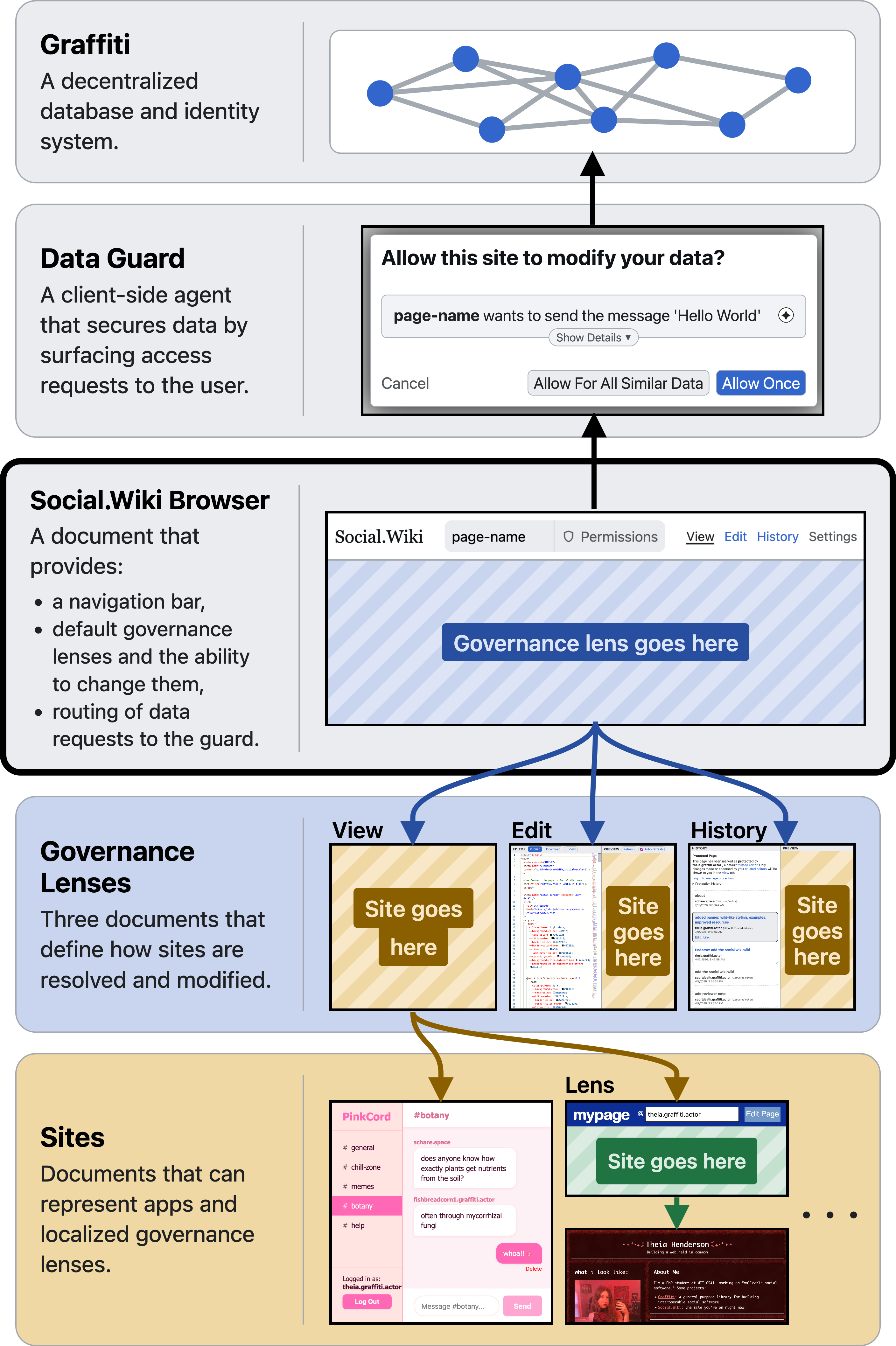}
    \Description[A system diagram of Social.Wiki with 5 layers, Graffiti, data guard, browser, governance lenses, sites]{
     The figure consists of 5 layers. The middle layer, 3, is bigger and bolded and says "Social.Wiki Browser". Arrows flow out from layer 3 both up and down. Going up are layers 2: Data guard, and 1: Graffiti. Going down there are layers 4: Governance Lenses and 5: Sites. Layer 3, "Social.Wiki Browser", has the description "A document that provides: a navigation bar, default governance lenses and the ability to change them, routing of data requests to the guard". Layer 3 contains an image that has a browser-bar-like header that says "Social.Wiki", then an address bar with the text "page-name", a button "Permissions", and four tabs View, Edit, History, and Settings. View is selected. Below the header is a blue striped area with the text "Governance lens goes here". Layer 2, "Data Guard", has the text "A client-side agent that secures data by surfacing access requests to the user". Layer 2 contains an image of a popup with the title "Allow this site to modify your data?", 3 buttons "Cancel", "Allow For All Similar Data", and "Allow Once", as well as a description "page-name wants to send the message Hello World" with an AI sparkle and a dropdown that says "Show Details". Layer 1, "Graffiti", has the description "A decentralized database and identity system." Layer 1 contains an image of an abstract decentralized network. There are arrows connecting the image in layer 3 to the image in layer 2 to the image in layer 1. Layer 4, "Governance Lenses" is blue and contains the description "Three documents that define how sites are resolved and modified." It contains three images labeled "View", "Edit", and "History". These images are connected by blue arrows to the blue striped area in Layer 3. The View image is a box with yellow stripes that says "Site Goes Here". The Edit image is a box with two panes including code on the left and a yellow striped box on the right that says "Site Goes Here". History contains two panes with the left containing different site version descriptions and the right containing a box that says "Site goes here". Layer 5, "Sites", is yellow and has the description "Documents that can represent apps and localized governance lenses." It contains two images connected to the "View" image in Layer 4 via yellow arrows. The left image is a screenshot of a "PinkCord", a pink chatting app. The right image is labeled "Lens" and has a header labeled "mypage" with the username "@theia.graffiti.garden" in a bar. Below the header is a green striped box labeled "site goes here." Within the site layer, there is an image linked to the lens image via a green arrow. The image is a MySpace-looking personal homepage. There are three dots following the sites.
    }
    \caption{The Social.Wiki system. Using Social.Wiki starts by loading its \emph{browser}. The browser connects to \emph{Graffiti}, a decentralized database and identity system, via an intermediate \emph{data guard} that confirms requests with the user.
    The browser also provides a navigation bar and three default \emph{governance lenses}: View, Edit, and History.
    The View lens queries the Graffiti database to resolve the name entered in the navigation bar into a document that is displayed as a \emph{site}.
    Sites may also query the database to populate themselves with posts, messages, and other user data.
    The browser provides a means of modifying the governance lenses, changing the governance of the system. Alternatively, sites can be used to resolve other sites, providing localized governance lenses.
    }
    \label{fig:system-diagram}
\end{figure}

A diagram of the system is shown in Figure~\ref{fig:system-diagram}.

One notable feature of Social.Wiki is that the only active servers involved in its operation are the generic Graffiti servers.
The rest of the system is a hierarchy of interactive \emph{documents}.
The browser is a document that uses Graffiti to dynamically fetch and transclude governance lens documents, which use Graffiti to dynamically fetch and transclude site documents. In other words:

\begin{quote}
\textbf{Social.Wiki} is a system of recursively transcluded interactive documents, where each document in the recursion is given guarded access to a shared database.
\end{quote}

We classify the documents that fetch and transclude other documents as ``lenses,'' a concept borrowed from functional programming~\cite{lenses}.
The governance lenses, as well as the browser, are lenses. 
Additionally,
sites---which can be just as expressive as the browser or governance lenses---can themselves be used as lenses, providing localized governance views, as later demonstrated in Section~\ref{sec:mypage}.

The browser
is currently distributed on the (world wide) web at \texttt{https://social.wiki} via the static hosting service GitHub Pages. 
However, it could just as easily be distributed over email or via a flash drive.
In the future, others might develop alternative browsers, just like alternative web browsers. 
So long as these competing browsers connect to the same database and make similar queries, they will interoperate with the original.
Therefore, at the ecosystem level, the browser is also modifiable, making the system plastic all the way through.

\subsection{Graffiti: Serving Sites and User Data}
\label{sec:graffiti}

Social.Wiki requires some sort of database to publish and query several types of content: the documents representing different proposed versions of a site; annotations of those documents to guide governance decisions, such as protection labels; and personal data displayed within sites themselves, such as messages and reactions.
We choose to use a system called Graffiti as this database.

Graffiti provides a general-purpose client-side JavaScript API for publishing and fetching data
with built-in design considerations specifically for \emph{social} data such as attribution, access control, and context management.
Additionally, the Graffiti API is implemented on top of a \emph{decentralized} architecture---people can freely choose the provider that hosts their own Graffiti data or even self-host~\cite{graffiti}.
Importantly, no specific Graffiti server is used to host any specific Social.Wiki site.
Instead, Social.Wiki sites are resolved by querying the \emph{network} of generic Graffiti servers as a large virtual database.

Graffiti differs from other decentralized social infrastructures such as the AT Protocol~\cite{atprotocol} (underlying Bluesky) and ActivityPub~\cite{activitypub} (underlying Mastodon)
because its client-side API is general enough that a wide variety of applications can be built on top of it with \emph{no additional servers}.
By comparison, building a custom application on the AT Protocol requires running a separate ``App View'' server, and building a custom application on ActivityPub requires hosting a separate server ``instance''~\cite{graffiti}.
By eliminating the need for additional servers, Graffiti allows a complete social application to be expressed within a single HTML document---the medium we choose for representing and editing sites on Social.Wiki\footnote{While the Hypertext Markup Language (HTML) was originally developed to represent text documents, the addition of CSS for styling and JavaScript for interactivity makes HTML a rich graphical and computational environment.
Webstrates~\cite{webstrates} similarly uses HTML as its medium, using synchronization servers for data persistence and sharing (see \S\ref{sec:malleable}), rather than Graffiti servers.
HTML also has the benefit of running on practically any device due to the ubiquity of web browsers.}.
This is especially important for Social.Wiki because it reframes ``shared ownership of a social application'' to be ``shared ownership of an HTML document,'' rather than ``shared ownership of a server.'' In doing so, we avoid the legal and physical complexities of jointly owning servers.

Graffiti data is managed \emph{individually}: a person can post and delete their own data, but cannot either create data on behalf of someone else or delete someone else's data.
While this may seem counterproductive to our goal of common ownership, clients can flexibly interpret collections of individual data objects as shared interactions.
For example, Social.Wiki sites are stored within Graffiti as a series of individually edited versions, but the View lens processes those edits to produce the experience of a collaboratively edited document.
\citet{graffiti} call this approach
``total reification'' as all interactions between people are reified into individual data objects.
Importantly, total reification makes it possible to apply different moderation policies to the same underlying set of reified user actions.
Social.Wiki's plural governance is implemented via this property by applying different selection algorithms to the various proposed versions of a site.

Graffiti data is organized into ``\emph{channels}.'' Like hashtags, channels are identified by arbitrary strings of text, do not need to be explicitly instantiated, and are \emph{not owned}---anyone can post to a channel.
A Social.Wiki site's name is used as the channel for broadcasting edit metadata for that site.
Importantly, it is not possible to find Graffiti data without explicitly querying one of its associated channels, a property designed to prevent ``context collapse''~\cite{contextcollapse}, which in the case of Social.Wiki allows sites to be effectively ``unlisted'': impossible to find unless given the exact name, especially if the name is long and random.

Graffiti provides Social.Wiki with substantial expressive power, allowing applications for messaging, microblogging, collaborative editing, and ridesharing to be represented entirely as HTML documents.
However, Graffiti applications cannot ordinarily be edited by simply clicking a button.
They are produced and governed just like other websites, 
arguably leading to the concentration of power outlined in Section~\ref{sec:motivation}.
Social.Wiki instead enables people to openly collaborate on Graffiti applications and to govern that collaboration.
Interestingly, Social.Wiki uses Graffiti to implement the lenses that facilitate collaboration and governance; however, this is not a straightforward feature of Graffiti but rather a novel use of Graffiti as a general-purpose database.
Finally, we introduce a data guard that secures access to Graffiti.
Although the data guard is particularly useful in Social.Wiki's dynamic and ownerless environment, it is also applicable to Graffiti applications more broadly, as discussed in Section~\ref{sec:dataguard}.

\subsection{Plural Governance}

The web is governed by a system of ownership.
In removing that ownership, we are faced with the difficult decision of choosing an alternative system of governance for resolving conflicts and maintaining quality.
While we do provide a specific \emph{default}, based on Wikipedia's governance, we propose the unusual approach of avoiding a concrete decision by allowing people to \emph{choose their own governance}.
This choice draws on pluralist approaches to design~\cite{escobarpluriverse,designjustice,feministhci} and on governance scholarship that rejects the idea that any single normative governance model will ever be appropriate for all communities and situations~\cite{beyondpanaceas,modularpolitics}.

To put our plural governance in perspective, some amount of personalized governance is already essential to most social platforms via features such as ``blocking,'' which affects one's own experience but not other people's.
However,
block lists are laborious to maintain and therefore unsustainable as the only form of governance~\cite{personalizingcontentmoderation,securingfederatedplatforms}.
Systems like SquadBox~\cite{squadbox} and Trustnet~\cite{trustnet} demonstrate how the labor of blocking people and posts can be reduced by delegating to close friends or other trusted sources, a feature now widely 
deployed in Bluesky under the name ``composable moderation''~\cite{composableModeration}, which allows one to subscribe to different moderation services.
Social.Wiki generalizes the personalization of governance beyond layers of block lists to the choice between different governance \emph{policies}, which may involve context-specific delegation, reputation systems, automation, and so on.

The crux of governing a web is the resolution of \emph{names}.
It is fairly trivial to create a system allowing anyone to publish (their versions of) documents under gibberish identifiers.
What is challenging is deciding which of those documents should be displayed when one visits a site with a human-readable name, like ``facebook''.
The World Wide Web takes the relatively straightforward approach of hierarchically assigning these names to owners via the Domain Name System (DNS)~\cite{dns} and having those owners choose how their names (URLs) are resolved\footnote{Decentralized alternatives to DNS like IPNS~\cite{ipfs}, Tor~\cite{tor}, and Namecoin~\cite{namecoin} still assign names to owners but via cryptography rather than a central arbiter.}.
In Social.Wiki, names are not owned. Instead, names are resolved independently by each user according to the policy described by their governance lenses.

To ground the governance possibilities in a concrete example, we first discuss how our default system of governance is implemented within the governance lenses. Then, we explore how those lenses can be modified to enable plural governance.
Further consequences of plural governance are discussed in
Section~\ref{sec:pluralreality}.

\subsubsection{Default Governance}

Our default governance assumes,
as do many peer production systems, that most participants act in ``good faith''~\cite{goodfaithcollaboration}, and therefore sites are openly editable by default (a ``do-ocracy''~\cite{doocracy,enablingcreativechaos}). However, changes can be reverted via the History tab.
In cases of prolonged disagreement  or ``edit wars''~\cite{editWars}, sites can be marked as ``protected'' by a ``trusted editor.'' When a site is protected, only edits made by a trusted editor are displayed. 

Even our default governance adds a degree of pluralism not found on Wikipedia: users may \emph{choose} who their trusted editors are, so a site can be protected for some people and not for others.
This property, inspired by the system Trustnet~\cite{trustnet}, is included because Social.Wiki is intended to contain many communities, and we do not expect any central body of administrators to know of every community's existence, let alone have the context to moderate them.
By allowing people to choose who they trust, the system has the potential to adapt to these small, disconnected communities.

The View lens implements this default by querying Graffiti for the logged-in user's trust relationships plus two kinds of information based on the name provided in the browser's navigation bar: published versions of the named site and any protection labels applied to it.
Using trust and protection data, the lens determines whether the site is protected for that user. If so, it excludes versions published by people the user does not trust. It then selects the most recently published eligible version and transcludes its associated HTML document to render the site. If no eligible version exists, the View lens displays ``Nothing here\ldots yet'' and links to the Edit lens.

Our default system of governance begs many questions:
What if I trust someone to edit certain sites but not others?~\cite{wikigrowsup}
Could the set of trusted editors be determined dynamically by a reputation system?~\cite{wikireputation}
Could AI be used to automatically block edits based on their content?~\cite{ores, potthast2008automatic}
These can all be addressed by modifying the governance lenses.

\subsubsection{Modifying Governance Lenses}

The governance lenses are implemented as HTML documents, just like the sites they govern.
The document used for each lens can be replaced within a Settings tab, provided by the Social.Wiki browser.
Alternatively, lenses can be replaced by clicking on a link that encodes the new documents, making the process of changing the lenses a one-click experience like installing a browser extension. 
To mitigate errors, the browser warns the user before making a change and provides an option to reset to the default lenses.

As all of the underlying Graffiti data is ``totally reified'' (\S\ref{sec:graffiti}), alternative View lenses can reinterpret the site versions, protection labels, and trust relationships provided by our default lens.
They can also query Graffiti for entirely new types of data---endorsements, flags, tags, \emph{etc.}---which alternative Edit and History lenses can be modified to produce.
One such alternative system might replace 
our default system of reactive site protection
with \emph{pro}active peer review where new edits to sites are not shown until
endorsed by at least two trusted editors.

Changes to the governance lenses are only applied locally, providing a base case in what would otherwise be an infinite nesting of governance layers.
However, there are cases where a user might want their governance updated on their behalf.
This can be achieved by installing a governance lens that introduces another layer of indirection via an additional lens.
Suppose, for example, that a user wants their Social.Wiki experience to follow rules chosen by their community's leaders. Their View lens could look up the rules most recently endorsed by those leaders and transclude the HTML document that implements them as a lens. That lens would then determine which site to display, producing the nesting ``governance lens $\rightarrow$ endorsed lens $\rightarrow$ site.''

We expect that most people will not develop their own governance lenses, instead choosing those created by others.
As governance lenses are just documents, they can be shared via direct message,
advertised in community spaces, or
published in a hypothetical ``governance store,'' similar to an app store.
Alternatively, a system like PolicyKit~\cite{policykit}
could be adapted to create lenses from higher-level primitives. None of these distribution or lens-crafting mechanisms require any new infrastructure to deploy---they can be built within Social.Wiki itself.

\subsection{Security}
\label{sec:security}

If Social.Wiki's governance were perfect, with all sites vetted by experts working in one's interest, then we would already be at the end of our design story.
Of course, as we have discussed, no governance will ever be perfect.
While we hope that most of the time the governance that people choose for Social.Wiki is good enough not to expose them to written or visual content that they would consider harmful, Social.Wiki involves the co-production of \emph{code} which introduces the possibility for small changes in that code to cause massive damage.

In particular, Social.Wiki requires dynamically executing other people's code \emph{and} giving that code access to read and write personal data.
Both of these present possible security vulnerabilities, which we address here.

\subsubsection{Sandboxed Transclusion}
\label{sec:sandbox}

Dynamically executing untrusted code is a classic security risk because, without proper protections, the running code can observe and modify its execution environment.
If this were the case within Social.Wiki, visiting a malicious site could inject code into the governance lenses or browser, which could in turn manipulate the governance, leak personal data, \emph{etc.}

Fortunately, modern browser engines are already excellent at isolating execution environments. This isolation is what makes it possible to visit a website without that site modifying either the browser or sites running in other tabs.
For our usage, browser engines also provide the ability to isolate transcluded documents: the \texttt{<iframe>} tag, which is used to include HTML within HTML, can be configured to run as a sandbox~\cite{mdn_iframe_sandbox},
completely isolating the transcluded code from its parent, except for a channel to pass messages between them.

Unfortunately, sandboxing leads to all sorts of restrictions, especially when sites do not have ``origins''---Social.Wiki sites are fetched from Graffiti and so are transcluded by their HTML source rather than via an HTTP link.
Most of our engineering efforts on this project involved wrestling with these restrictions.
Functionality as simple as following links between sites has been reconstructed by bubbling navigation messages up through the iframe tree.

Every Social.Wiki document (browser, governance lens, or site) includes a line of code that injects the script \url{https://social.wiki/init.js}. This script establishes a message connection to the document's parent and provides the web component \texttt{<sw-transclude>}, which can be used to transclude other documents. \texttt{<sw-transclude>} does the work of automatically configuring a sandboxed \texttt{<iframe>} and establishing a message connection to it.

\subsubsection{Data Guard}
\label{sec:dataguard}

Graffiti is used within Social.Wiki to publish sites, trust editors, send and receive private messages, and perform all other social actions within the system.
Graffiti data is also interoperable, ensuring that social interactions persist across different versions and forks of a site and different governance systems. 
However, all of this expressivity and interconnectedness creates risk.
Without proper protection,
a malicious site could delete all of a person's Graffiti data, publish forged messages on their behalf, leak private messages, and so on.

To mitigate this risk, sites are not given \emph{direct} access to Graffiti at all.
Instead, only the ``data guard,'' embedded in the browser, is given custody of the Graffiti credentials needed to modify data and read private data.
To perform a Graffiti operation, a site sends a request to the browser, which in turn forwards the request to the guard, which confirms the request with the user via a pop-up, as shown in Figure~\ref{fig:teaser}, panel 3.
The user is presented with an approachable summary of the request, generated by a local AI model~\cite{webllm}, and can also inspect the exact request details. If the user allows the operation, the result is returned to the site.

Effectively, the data guard is guarding access to Graffiti, just like a web browser guards access to one's camera or microphone.
This granular security model, based on the Principle of Least Privilege~\cite{saltzer1975protection}, allows one to safely browse and even interact with random Social.Wiki sites, with the certainty that the sites will not modify or access any data other than the data they explicitly request and are granted access to.

Of course, for continued use of a site, these constant pop-ups will surely induce ``confirmation fatigue.''
To mitigate this, the pop-ups give the option to ``Allow For All Similar Data.''
This option grants the requesting site continued permissions to perform ``similar'' data operations until revoked or the site is changed\footnote{Note that the permissions system does not explicitly know that a site is changed because site changes are defined by the View lens. Instead, the permissions system tracks transcluded documents by IDs that are assigned recursively from parent to child documents. The View lens assigns IDs to its children based on the hash of their HTML documents, which makes a site's ID change whenever the site's content is changed.}.
Permissions can be revoked via the Permissions tab that appears in the browser's navigation bar (visible in the browser in Figure~\ref{fig:system-diagram}).

Currently, data similarity is assessed via a deterministic set of rules that match data objects with the same property names and value types.
However, in the future, a local AI agent could offer a more nuanced notion of similarity, perhaps even detecting messages that seem uncharacteristic for the sender.

For people developing Social.Wiki sites, the complexity of passing Graffiti requests to the data guard is abstracted away.
The \texttt{init.js} script described in Section~\ref{sec:sandbox} exposes a global class called \texttt{window.Graffiti}.
This class provides all the methods of a normal Graffiti implementation, only under the hood these methods actually send messages to the data guard.

The data guard is reusable in Graffiti apps outside of Social.Wiki.
To achieve this, the guard's code is hosted on its own domain and embedded into the target app, \emph{e.g.}, our browser, via an iframe. 
Anti-cross-origin-scripting features ensure that the embedding app cannot access the guard's credentials. Effectively, the guard acts as an external server, even though its code runs entirely client-side.
\section{Deployment}
\label{sec:deployment}

We gradually deployed Social.Wiki to friends and colleagues over two months and then hosted a publicly advertised hackathon. At the time of writing and based on the information available (it is a decentralized system) Social.Wiki has had at least 100 unique users.

To reduce the barrier to entry,
the Social.Wiki landing page links to a guide that walks through ``vibe coding'' a Social.Wiki site.
This involves copying our meta-prompt describing the system into an AI chatbot or coding agent.
In a single shot, these systems can generate messaging apps, dating apps, board games, collaborative canvases, and so on.
While vibe coding can introduce security risks if used improperly, sites are sandboxed and requests to modify user data are surfaced to the user (\S\ref{sec:security}). This makes it safe to interact with a site built by any amateur programmer, including a coding agent.
Many, but not all, sites were at least partially vibe coded.

The sites that have received the most continued use so far are those that are in support of existing community groups, specifically a reading group and a film club.
These sites both began as static sites with their schedules directly embedded in the HTML. However, over time---due to the efforts of multiple editors---each accrued social features including comments on each reading and a collectively curated watch list.
Effectively, these sites evolved into ``microplatforms'' in direct response to user needs, and they can continue growing and evolving as they are used. There is no ``webmaster'' who is on the hook to implement desired changes or who has the privilege of dictating the final design.

Given that Social.Wiki usage is currently relatively small, we have not seen the full use of its governance affordances, which are instead evaluated through case studies (\S\ref{sec:case-studies}).
However, we have observed two lighthearted disagreements.
First, the film club site underwent a brief edit war when a user registered an anonymous Graffiti identity and changed the watch schedule to set \emph{Twilight} as the next movie, triggering amusement in the film club's group chat. Second, a chat app for passing notes in class, named simply ``chat,'' was forked to create a new app called ``betterchat'' that altered the interaction design so users could send multiple notes at a time.

\subsection{Hackathon}
\label{sec:hackathon}

Our hackathon was a three-hour event that attracted 12 participants. After a brief introduction to the system, people spent two hours building before presenting their work and filling out a survey.

At the end, three awards were given out: ``Most Likely To Be Used,'' to an internal housing co-op communication tool; ``Most Creative,'' to a microblogging site where posts are planted as flowers in a garden; and ``Trust and Safety,'' to an end-to-end encrypted Slack clone\footnote{Note that Graffiti is not currently end-to-end encrypted, but there is ongoing work to do so~\cite{graffiti}, which would make all Social.Wiki sites end-to-end encrypted by default.}.
Other highlights include a collaborative Python editor\footnote{Python can be executed in the browser using Pyodide.}, a choose-your-own-adventure game where users can collaboratively add branches, a Social.Wiki search engine, a dating app for polycules, and a version of Conway's Game of Life that is seeded by data posted to other sites.
Some examples are shown in Figure~\ref{fig:teaser}.

11 out of 12 participants completed an exit survey.
All self-reported some programming (median: 4/5, min: 2, max: 5) and front-end experience (median: 3/5, min: 2, max: 5).
Future work could evaluate Social.Wiki for non-experts (although, due to AI, most participants were not ``programming'' in a traditional sense).

\subsubsection{Developer Experience} A number of participants spoke positively of the development experience (P2, P4, P5, P7, P8, P9, P10), with P5 noting it was ``Way easier and way faster than any other development experience I've ever had'' and P8 noting ``it felt powerful functional and value aligned!'' However, there was also room for improvement: P2 and P10 wanted an improved editor, P4 and P1 wanted the ability to split development over multiple files, and P8 wanted better debugging.
For the sake of time, participants were not told about the extent of Social.Wiki's plasticity, which in theory could be used to address these issues.

Participants appreciated how batteries-included the system was (P3, P7, P8, P10, P11); \emph{e.g.}, P8 liked ``The rapid ability to prototype and not having to spend any time worrying about deployment, data models, servers''.
P9 expressed a wish for more documentation and ``Having trouble figuring out how to start my project.''
However, P7 pointed out, ``discoverability is simple and fast enough that someone can figure it out just by clicking around.''
Finally, P5 and P9 expressed desires for a ``directory that lets me see what's out there and how pages relate to each other!'' (P9). Fortunately, P11 built a social search engine (Fig.~\ref{fig:teaser}, d) that does exactly that!

\subsubsection{Collaboration} Most hackathon participants decided to work alone and so did not take advantage of collaborative features. However, P11 noted, ``It was cool to easily create a website that could be used by anyone and could be kept up even if not by me.'' Despite editing alone, participants were observed regularly visiting and interacting with each other's sites throughout the hackathon.

\subsubsection{Permissions}
Permissions were a noted pain point for P7, who compared them to ``disallowing cookies on every site.''
However, P5 said, ``I love the feature of being able to see what you're uploading (the "allow" button)'' and wanted even more transparency on data being shared.
This suggests that future work should provide means of delegating or automating more of these security decisions, similar to the cookie Consent-O-Matic system~\cite{consentomatic}, while maintaining the granular control and transparency for others who want them. 

\subsubsection{AI} Participants were split over AI.
Some thought that it made development easier (P2, P6, P7), and P5 liked using a chatbot primed with our prompt as a tutor to learn about Graffiti and Social.Wiki. However, P1 found it frustrating and P6 felt it did not allow them to understand the ``inner workings'' of their application.

AI is not inherent to using Social.Wiki but was encouraged during the hackathon to ensure that people with a range of skills could build something interesting in a short amount of time.
It is an open question as to what alternative low/no-code tools might be created for development and what role AI will play in the sites intended for long-term use rather than rapid prototypes.
\section{Case Studies}
\label{sec:case-studies}

\begin{figure}[htbp]
    \centering
    \setlength{\fboxsep}{0pt}%

    \begin{subfigure}[b]{0.48\columnwidth}
        \centering
        \fcolorbox{black}{white}{%
            \includegraphics[width=\textwidth,keepaspectratio]{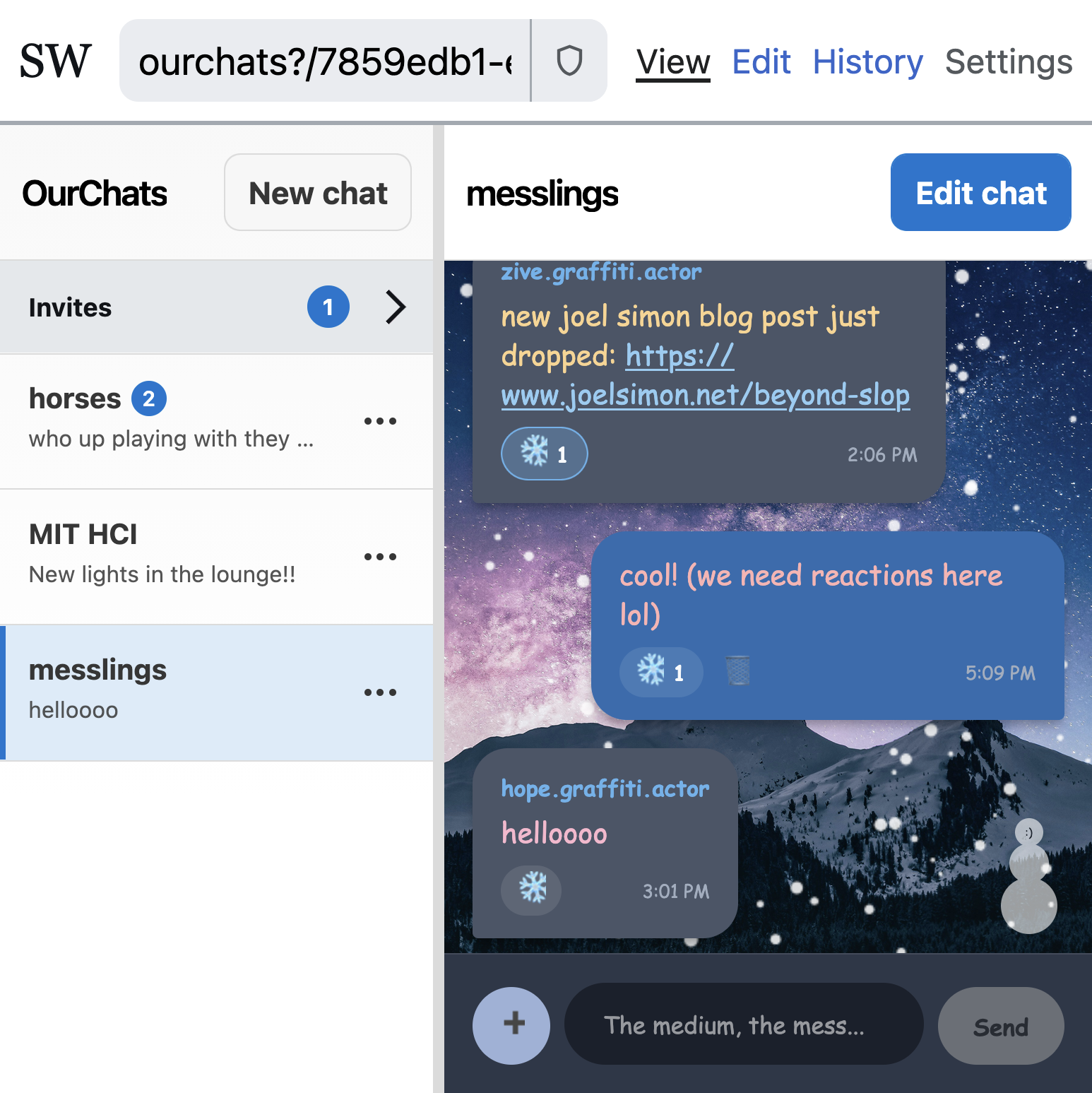}%
        }
        \caption{OurChats}
        \label{fig:ourchats}
    \end{subfigure}
    \hfill
    \begin{subfigure}[b]{0.48\columnwidth}
        \centering
        \fcolorbox{black}{white}{%
            \includegraphics[width=\textwidth,keepaspectratio]{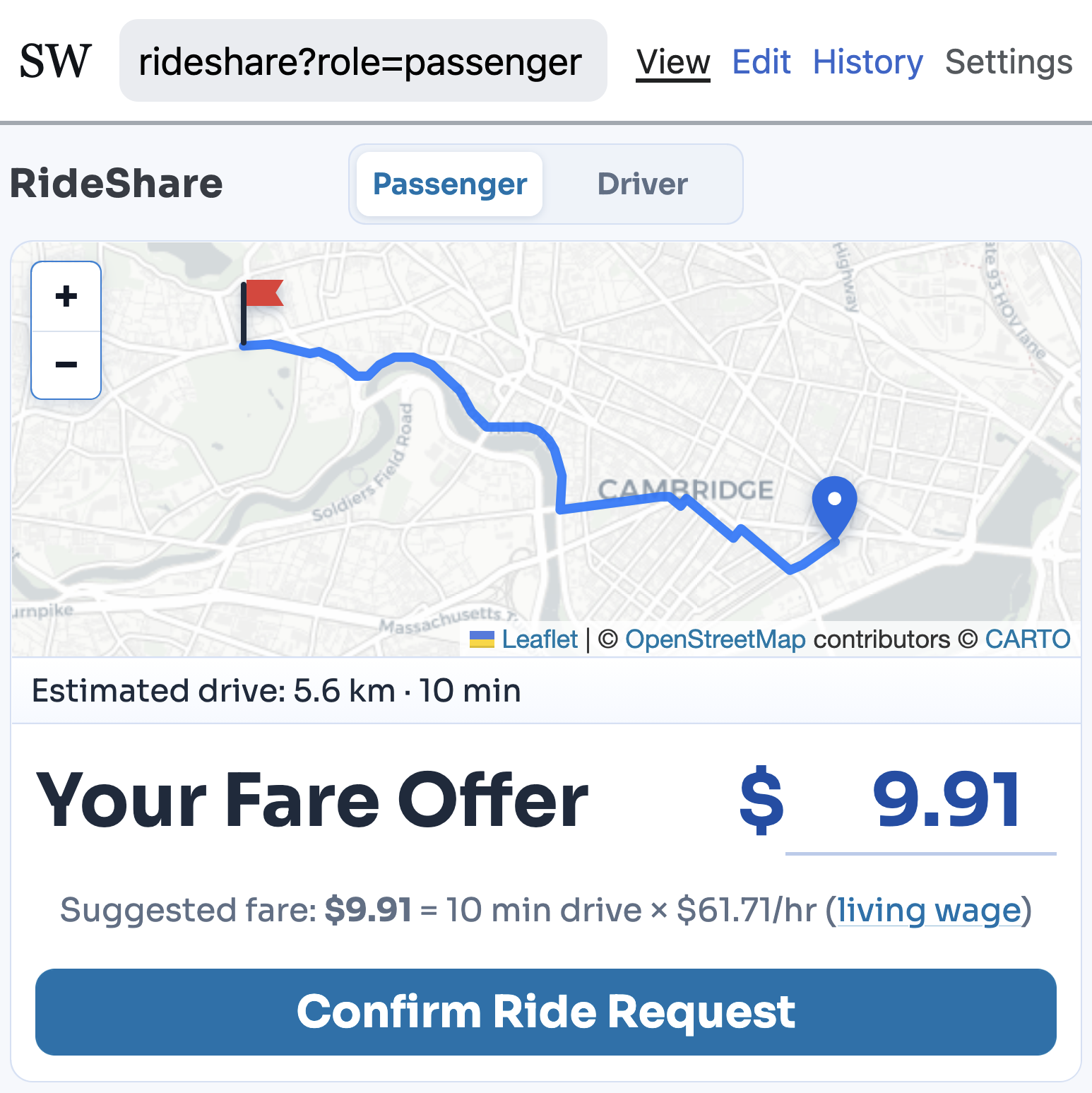}%
        }
        \caption{Rideshare}
        \label{fig:rideshare}
    \end{subfigure}

    \par\addvspace{1em}

    \begin{subfigure}[b]{\columnwidth}
        \centering
        \fcolorbox{black}{white}{%
            \includegraphics[width=0.8\textwidth,keepaspectratio]{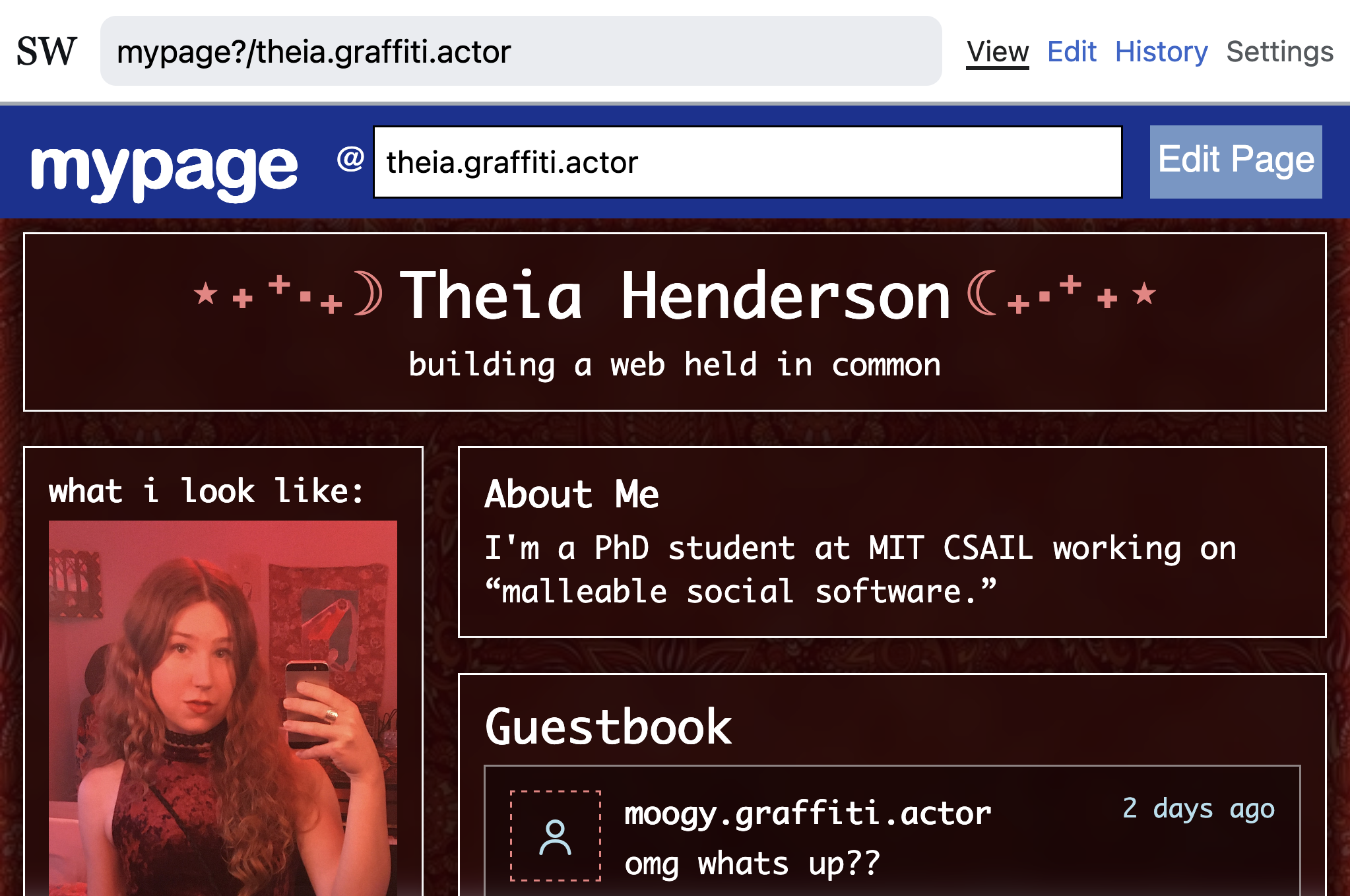}%
        }
        \caption{Mypage}
        \label{fig:mypage}
    \end{subfigure}

    \caption{Screenshots from our case study sites.}
    \label{fig:combined}
    \Description[A messaging app with a customized chat UI, a ridesharing app, and a myspace-like homepage] {
    The figure is divided into three subfigures, a, b, and c.
    Subfigure a contains a screenshot of a browser-like interface with a chat app displayed. The chat interface itself is highly customized. The top bar of the browser says "SW" and has the text in its navigation bar, "ourchats?/" followed by a random ID. The bar has the buttons "View", "Edit", "History", "Settings" to the right. The chat app has two panes, a sidebar on the left and a chat interface on the right. The sidebar says the title "OurChats", an ability to create a new chat, an invite list with notifications, and three chats with notifications and most recent messages displayed. The chat interface on the right has a bar at the top naming the chat with option to "Edit chat". The chat itself has a snowy interface with colorful Comic Sans messages and snowman reactions.
    Subfigure b contains a screenshot of a browser-like interface with a ridesharing app displayed. The top bar says "SW" and has the text "rideshare?role=passenger" in the navigation bar with the buttons "View", "Edit", "History", "Settings" to the right. The app itself has a toggle between passenger and driver with "Passenger" selected. There is a map with a route in blue with a start and end point and "Estimated drive 5.6km or 10 minutes". Below the map there is "Your fare offer" with \$9.91 entered. Below there is the text "Suggested fare: \$9.91 = 10 minute drive x \$61.71/hr (living wage)". Finally, there is the text "Confirm Ride Request".
    Subfigure c contains a screenshot of a browser-like interface containing a Myspace-like personal homepage. The top bar says "SW", has the text "mypage?/theia.graffiti.actor" in the navigation bar and the buttons "View", "Edit", "History", "Settings" with View selected. Below the top bar there is another bar that says "mypage" its navigation bar says "@theia.graffiti.actor" and to the right there is an option to "Edit page". Below there is a personal homepage containing a stylized website with the title "Theia Henderson" an "about me" section, selfie. There is a guestbook section with one entry from "moogy.graffiti.actor" visible.
    }
\end{figure}

\subsection{OurChats: A Malleable Messaging Platform}
\label{sec:ourchats}

OurChats is a messaging platform, akin to Facebook Messenger. 
What makes OurChats unusual is that each chat includes an “Edit” button that allows participants to modify that particular chat's styling and interface features. For example, one OurChats chat has been modified to use rainbow Comic Sans font, have an animated snowy background, and include snow reactions, as shown in Figure~\ref{fig:ourchats}.

Under the hood, any time a new chat is created, it is actually creating a new \emph{site} from a basic chat template.
This site is transcluded within the OurChats site so that it can be displayed along with the OurChats sidebar that lists all chats.

OurChats demonstrates how the benefits of malleability can be integrated into an otherwise conventional chat interface---there is no need to manually create a new site from scratch just to form a new conversation and no need to manually check a variety of independent chat sites for new messages.
This per-chat malleability allows people to bring the features that they like to express themselves with~\cite{Griggio_2019} to any chat,
without creating a complex super-app that has all the possible features of every chat.

Additionally, OurChats demonstrates how  transclusion can be used to delineate governance boundaries.
Regardless of whether OurChats is protected (or otherwise governed under other systems), the fact that the chats within OurChats are separate sites means that they are governed independently.

Subtly, transclusion within OurChats is used differently than transclusion in the View lens. The View lens runs a resolution algorithm (\emph{e.g.}, as in Section~\ref{sec:default}) that transcludes a document by \emph{source}.
OurChats does not reimplement that resolution logic.
Instead, it transcludes by \emph{reference}, bootstrapping off of the user's chosen View lens rather than introducing new governance.

\subsection{Mypage: A Page as a Governance Lens}
\label{sec:mypage}

Mypage is a site for making and browsing personal homepages, similar to Myspace or Neocities.
You can visit someone's ``Mypage page'' by entering their Graffiti username into a navigation bar.
A Mypage page can include both static content and social features, like the guestbook in Fig.~\ref{fig:mypage}.
Importantly, unlike the rest of Social.Wiki, inside Mypage \emph{you cannot edit someone else's Mypage page}. 

Under the hood, Mypage is a lens that works much like the View lens.
It treats a username entered in the Mypage bar as a site name.
Then, it uses Graffiti to look up edits associated with that name and transcludes the document corresponding to the most recent edit.
However, unlike the View lens, Mypage only considers edits to a site by \emph{one person}---the person whose username was entered.

Mypage produces a governance of \emph{ownership}.
While we began this paper critiquing the ownership of \emph{shared social spaces}, there is of course still value in having ownership over one's personal identity, as clarified in Section~\ref{sec:personaldata}.
Mypage allows one to express their personal identity through an interactive webpage.
However,
if Mypage were used to create a community platform and there was a falling-out between community members and the creator, members would not be forced to experience the creator's ``personal expression.''
Instead, they could remove the Mypage lens and view the page within Social.Wiki as a whole, where it is openly editable.

Mypage also demonstrates a routing convention respected by the default View lens and other lenses. 
In the name `\texttt{mypage?/user},'
the text `\texttt{/user}', following the special character `?', is not treated as part of the site name but instead is sent as a parameter to the transcluded site.
This makes it possible to directly link to a particular page within a site, similar to client-side routing in single-page apps.

Finally, note that even though Mypage does not allow people to edit each other's Mypage pages, anyone can edit Mypage itself. Therefore, if Mypage were to become popular, it might need to be governed (via protection or otherwise) to prevent people from ``breaking'' Mypage's governance, which could (among many other possibilities) be used to vandalize Mypage pages.
Of course, ``breaking'' is subjective here as there is no reason why the name Mypage \emph{must} be associated with this particular functionality.

\subsection{A Collectively Owned Ride-Sharing App}
\label{sec:rideshare}

``Rideshare'' is a Social.Wiki site for matching riders to drivers, much like Uber or Lyft (see Fig.~\ref{fig:rideshare}).
A rider selects where they want to be driven from and to.
Then, the rider is presented with a suggested fare based on a transparent formula that multiplies the expected time it will take to drive along the route by a local ``living wage,'' per the Living Wage Calculator~\cite{livingwage}.
The rider can adjust the fare they want to pay for the ride before submitting a request.

In a separate pane, drivers can browse for nearby ride requests. The exact pickup and drop-off locations of the requests are not revealed to the drivers for privacy purposes, but drivers can see approximate locations as well as the fares being offered.
Once a driver accepts a given request, the rider is asked to confirm the driver, at which point the exact pickup and drop-off locations are revealed to the driver and the trip can commence.
Like a traditional taxi service, fare is expected to be paid directly by the rider to the driver at the end of the ride, via cash or digital payment.

Rideshare uses the peer-produced map service OpenStreetMap~\cite{haklay2008openstreetmap} for displaying interactive maps and calculating routes between locations.
To efficiently query ride requests in a given location, ride requests are indexed spatially into Graffiti channels according to an open-source global grid~\cite{h3} (interestingly, developed by Uber).

Rideshare demonstrates how Social.Wiki can be used to produce a multi-stakeholder cooperative that is accountable to both the riders and drivers.
While we can only speculate about how the competing interests will play out, it could result in more transparent matching algorithms, as well as adaptable policies for safety, dispute resolution, and working conditions.
Additionally, Rideshare does not extract any commission because it has no owners.
Of course, Rideshare does not solve all of the problems of the gig economy.
In particular, its decentralized nature means it cannot be used to offer a minimum wage or health insurance.
However, arguably these services might be better provided directly by public-benefit organizations like governments, unions, or mutual aid groups, rather than by employers who must be coerced into doing so.

So far, we have used Rideshare to successfully complete one ride.
The ride was performed voluntarily rather than for pay because it is not clear how Rideshare fits into existing local regulation: the companies that \emph{own} ride sharing platforms must be licensed, but what does that mean for a platform that is not owned?

For realistic use, Rideshare will of course need to be upgraded to include safety features such as driver and passenger reputation services and the ability to opt into different background check services, both of which are more than possible within Social.Wiki.

One technical note is that the sandboxing introduced in Section~\ref{sec:sandbox} does not allow Social.Wiki sites to access one's location---in Rideshare, one's starting location is provided manually.
In the future, the data guard used to grant access to Graffiti could be expanded to match the browser mechanism that inspired it by guarding other inputs like one's location, camera, or microphone.

\section{Discussion: A Plural Web}
\label{sec:pluralreality}

Plural governance makes it possible for different people to see
different versions of the same site.
From an individual perspective, this empowers people to shape their experiences to their own needs and preferences.
At the ecosystem level, however, this
plural governance raises several natural concerns: it might
cause confusion, diffuse collaborative effort, or erode the effectiveness of governance as a whole.
In this section, we argue that the divergence that plural governance makes possible will be minimal, as Social.Wiki provides many intermediate opportunities
to accommodate disagreement.
Furthermore,  we argue that, even when divergence occurs,
confusion and diffusion can be mitigated and governance is not made ineffective.

As a caveat, these arguments have not been evaluated at scale.
As we have deployed Social.Wiki to only roughly a hundred
participants (see  Section~\ref{sec:deployment}), we have
fortunately not seen cases where even our default page-protection
system was necessary, let alone cases where governance diverged.
Scaling Social.Wiki to the point where these mechanisms become
relevant must be done cautiously, by cultivating cultures of
good-faith collaboration within communities
~\cite{goodfaithcollaboration,teaandsympathy} and refining our
reference browser's default governance to provide an effective
baseline.

\subsection{Accommodating Difference}

Governance in general is a means of last resort when disagreement cannot
otherwise be accommodated.
As Social.Wiki sites are interactive, interoperable software, they can accommodate more difference than if they were simple static articles.
This creates a spectrum of intermediate forms of divergence before conflicting governance policies sever people's views completely.

One approach is to turn a point of disagreement into a
within-site \emph{setting}.
For example, a site does not strictly need to be in either light mode
or dark mode: it can allow each person to toggle between the two.
A user's personal settings may be stored, like any other data, on
Graffiti.
If there is controversy over the default settings, the site can
instead present newcomers with a setup wizard containing multiple
options.

If attempts at compromise fail, despite settings or traditional
good-faith collaboration strategies
~\cite{goodfaithcollaboration,boldrevertdiscuss}, one option short
of divergent governance is to ``fork'' a site, \emph{i.e.}, copy its
source code under a new name.
Since social data is stored on Graffiti rather than within a site,
competing forks can continue to interoperate with the same data even
as their code diverges.
Unfortunately, a fork no longer automatically benefits from
contributions to the original, potentially diffusing collaborative
energy, similar to platform exit as discussed in
Section~\ref{sec:motivation} and remix systems as discussed in
Section~\ref{sec:peerproduction}.
This diffusion can be reduced by decomposing a site into common
components shared between competing versions through transclusion,
a form of modularity often used in Webstrates~\cite{webstrates}.

If there is disagreement over who should keep the name of a forked
site, one amicable solution is for \emph{no one} to keep it and
instead replace the site at the original name with a disambiguation
page linking to the competing versions.
This is similar to how Wikipedia uses disambiguation pages for words
with multiple meanings.
However, the flexibility of Social.Wiki means that a disambiguation
page need not be a simple list of links; it could instead be
interactive, like a BuzzFeed-style quiz: ``Which site is right for
you?''

\subsection{When Governance Diverges}

If disagreement is so strong that people cannot even agree on a
disambiguation page, governance must finally be used to choose a
side.
Different governance systems may make different choices, causing
the same name to resolve to different site versions for different
people.
This divergence may make it confusing to discuss a site, since its
name no longer refers to the same content for everyone.
In such cases, people can link directly to a particular site
\emph{version}---each version automatically receives a long,
hash-based identifier---rather than relying on a contested
human-readable name.
Alternatively, one's View lens could surface ambiguities with a
warning such as ``Other people see different content than you'' and
a toggle for switching to a competing version.
People can also share their governance lenses as sites, just like
Mypage (\S\ref{sec:mypage}), allowing others to temporarily view
Social.Wiki through their eyes.

We conjecture that differences among governance systems will often
resemble those between the ``nightly'' and ``stable'' releases of
open-source software.
Some systems may receive updates almost immediately but occasionally
encounter broken or problematic content, while others may update
more slowly after changes have been vetted through various
processes.
Systems that otherwise present completely disconnected realities give up the
benefits of collaboration, creating pressure to converge where
possible.

However, even if governance policies diverge completely,
from an \emph{individual}'s perspective, their chosen governance policy
continues to work exactly as intended, protecting them from harmful
or unwanted content even if it exists elsewhere.
Social.Wiki can reproduce both Wikipedia-like protection and
traditional web-style ownership; pluralism does not make either of these highly scalable models any less effective for those who
choose them.

\subsection{Governing Reach, Not Existence}

Aside from external laws that make it illegal to distribute certain types of content, Social.Wiki's decentralized design, governance, and infrastructure make it largely impossible to prevent harmful content from existing somewhere in the system.
While this may seem alarming, it is no different from the existing web where it is almost always possible to host such content on an obscure domain.

This does not mean that Social.Wiki is neutral toward harmful content. Rather than pretending that such content can be made impossible to publish, Social.Wiki democratizes the design of content \emph{reach}: communities can determine how and what they endorse, flag, and help others discover. We believe that distributing this power can foster healthier ecosystems of community care that discourage harmful content from being produced and spread in the first place. We cannot guarantee that the results will always be fair or effective, as no system can, but instead give the people affected by governance a role in designing and revising it, rather than placing that authority in the hands of an arbitrary external owner.
\section{Conclusion and Future Work}
\label{sec:conclusion}

The ownership of websites made sense in the web's early days when it was largely used to browse personal homepages. However, as web traffic increasingly concentrates on a handful of dominant platforms, 
the owners of those platforms are ``becoming owners of the infrastructures of society''~\cite{srnicek2016platform}.
By combining ideas from wiki-style peer production and collaborative malleable software systems, Social.Wiki imagines an alternative web
where people can collectively reshape both the platforms they inhabit and the policies governing that collaboration.

While Social.Wiki is a functional and deployed system, our base of a hundred or so users is tiny compared to the existing web. Therefore, we have yet to see whether its system design or model of plural governance works at scale.
Future work is also needed to
refine our default governance policy
(many people never change defaults~\cite{mackay1991triggers});
relax Social.Wiki's exclusively client-side architecture so that it can support computation-intensive tasks like algorithmic ``For You'' feeds or sites with millions of edits;
improve the developer experience beyond single-file HTML editing for large, long-lived projects;
and foster a culture of good-faith collaboration.

Social.Wiki does not eliminate conflict or the need for governance.
Instead, it changes who has the power to respond to these challenges. Rather than leaving people as tenants within software designed and governed by distant owners, Social.Wiki gives communities a substrate for constructing, contesting, and continually revising their shared online spaces.

\begin{acks}
We would like to thank all of the hackathon attendees, with a special mention to Barish Namazov, Andre Ye, and Ryan Yen, whose creations are displayed in Figure~\ref{fig:teaser}.
Additionally, the first author would like to thank Catherine D'Ignazio for her class on ``Cities Without Property,'' which inspired this work; Benjamin Mako Hill for insights on peer production; and Helena Vasconcelos for figure guidance.
A. Dodik acknowledges the generous support from Schmidt Sciences and from National Science Foundation grant IIS2335492.
\end{acks}
\bibliography{bibliography}

\end{document}